\documentstyle[epsf]{l-aa}
\begin{document}
\def \hi {H\,{\sc i~}}
\def \hii {H\,{\sc {ii}~}}
\thesaurus{04(11.09.1 Dw1, 11.09.1 Dw2, 11.09.03, 11.11.1, 11.12.1,
11.19.2)}
\title{Neutral hydrogen in the nearby galaxies Dwingeloo 1 and
Dwingeloo 2}
\author{W.B. Burton \inst{1} \and M.A.W. Verheijen \inst{2} \and
R.C. Kraan--Korteweg$^{2,}$ \inst{3} \and P.A. Henning \inst{4}}
\institute{Sterrewacht Leiden, Postbus 9513, 2300 RA, Leiden, The
Netherlands
\and Kapteyn Astronomical Institute, Postbus 800, 9700 AV, Groningen,
The Netherlands
\and Observatoire de Paris--Meudon, DAEC, 5 Place Jules Janssen,
F-92195 Meudon--Cedex, France
\and Department of Physics and Astronomy, University of New Mexico,
Albuquerque, NM 87801, U.S.A.}
\offprints{W.B. Burton}
\date{Received \dots, 1995; accepted \dots, 1995}
\maketitle

\begin{abstract} We present observations made with the Westerbork
Synthesis Radio Telescope of \hi emission from Dwingeloo 1, a nearby
barred spiral discovered during the Dwingeloo Obscured Galaxies Survey
for galaxies hidden in the Zone of Avoidance, and of Dwingeloo 2, a
small galaxy discovered in the beam of these WSRT observations.  The
WSRT data reveal the position of the dynamical center of Dw1, its
systemic {\it LSR} velocity, its total (projected) width in velocity,
its inclination on the sky, and its integrated \hi flux, as well as
details of the velocity field and gas distribution.  Dw1 is the
nearest grand-design barred spiral system, and is probably amongst the
ten largest galaxies closer than about 5 Mpc.  We report here also the
discovery of Dwingeloo 2, a small galaxy located within the WSRT
primary beam as pointed to Dw1.  In view of its angular and kinematic
proximity to Dw1, Dw2 may well be a companion to the larger system.
The two galaxies are probably both members of the group containing
Maffei 1 \& 2 and IC342 and may influence the peculiar motions within
that group and the morphology of its individual members.

\end{abstract}

\section{Introduction; the Dwingeloo Obscured Galaxies Survey}

     Some 20\% of the optical sky is significantly obscured by
absorption and scattering of light by intervening interstellar
material in the Milky Way.  As a consequence of this obscuration, the
inventory of nearby galaxies is certainly incomplete closer than about
$5\degr$ to the galactic equator, where the optical and infrared sky
is grossly confused by local stars and other objects and where deep
optical searches for obscured galaxies fail (see Kraan--Korteweg \&
Woudt, 1994).  The boundaries of the Zone of Avoidance are difficult
to quantify; there are patches of sky beyond $|b| \sim 5\degr$ where
the optical depth is also substantial.  The intervening interstellar
medium which is largely opaque at optical wavelengths is transparent
at the 21-cm wavelength of neutral hydrogen.  Reviews of the use of
the \hi line to search for galaxies in the Zone of Avoidance have been
given by Kerr and Henning (1987), by Kerr (1994), and by Henning
(1992, 1994).

     The Dwingeloo Obscured Galaxies Survey has been searching for \hi
emission from galaxies hidden behind the obscuring material causing
the Zone of Avoidance, using the 25-m radio telescope of the
Netherlands Foundation for Research in Astronomy, located in
Dwingeloo.  The project is motivated by the hope to find the \hi
signatures of fairly nearby galaxies in the velocity range $0-4000$ km
s$^{-1}$.  Such detections would contribute to investigations of the
dynamics of the Local Group, of the morphological distribution of
nearby galaxies, of the age of the Universe as determined by the
timing argument, and of the density parameter $\Omega$ as revealed by
the velocity of the Local Group relative to the Microwave Background.

\begin{figure*}[t] 
\caption[]{Mosaic of channel maps showing the \hi emission from
Dwingeloo 1, after subtraction of the 21-cm continuum radiation and
after application of the {\it CLEAN} algorithm, but uncorrected for
primary-beam attenuation.  For each panel the appropriate central {\it
LSR} velocity is indicated; the cross indicates the position of the
dynamical center.  The ellipse in the upper-lefthand panel indicates
the approximate size and orientation of the optical image obtained
with the Isaac Newton Telescope in the $I$-band image (see Loan et al.
1995).  Contour lines in the channel maps correspond to $-$0.42,
$-$0.21, 0.21 (the 2-$\sigma$ level), 0.42, 0.85, 1.3, 1.7, 2.5, 3.4,
and 4.2 $\times 10^{20}$ \hi atoms cm$^{-2}$.  The first five channel
maps show contaminating \hi emission from our own Galaxy.  The
lower-righthand panel shows the (un-{\it CLEAN}ed) continuum radiation
from the field, for which the individual channel maps have been
corrected.  Contour lines in this continuum map correspond to $-$1.3,
$-$0.62, 0.62 (the 2-$\sigma$ level), 1.3, 2.5, 5.0, 7.5, 10, 13, and
15 K; dashed-line contours represent the negative values.}
\end{figure*}

     It is expected that the Dwingeloo Obscured Galaxies Survey will
continue, utilizing the telescope essentially full-time, for a
duration of several years.  Before this search began, the Dwingeloo
25-m telescope had been used over a five-year period to produce the
Leiden/Dwingeloo atlas of \hi in our Galaxy over the entire sky at
$\delta > -30\degr$ and within the velocity range $-450<v<+400$ km
s$^{-1}$ (Hartmann, 1994; Hartmann \& Burton, 1995).  That survey
represented the first use of the 25-m telescope with the new {\it DAS}
1000-channel digital autocorrelator spectrometer combined with the
receiver, characterized by a system temperature of about 35 K, which
is a prototype of the sort mounted on the telescopes of the Westerbork
array.  The experience gained with the Leiden/Dwingeloo Milky Way
survey confirmed that the interference environment was benign enough,
and the telescope electronics stable and sensitive enough, to justify
the search for obscured, relatively nearby, galaxies.

     The performance and operating configuration of the Dwingeloo
25-m telescope currently pertaining is that described by Hartmann
(1994), Burton \& Hartmann (1994), and Hartmann \& Burton (1995).  For
the obscured-galaxies project, the radio receiver is tuned to a
central frequency corresponding to a velocity of $+2000$ km s$^{-1}$
above the rest frequency of the \hi hyperfine transition; the
bandwidth of the 1000-channel spectrometer is set to cover the range
$0<v<+4000$ km s$^{-1}$ at 4 km s$^{-1}$ resolution.  (Radial
velocities are defined here following the radio convention and are
referred to the Local Standard of Rest defined by the Standard Solar
Motion of 20 km s$^{-1}$ towards $\alpha,\delta = 18^h,+30\degr$,
epoch 1900.)

     The Dwingeloo Obscured Galaxies Survey involves a strategy with
several different facets.  Two blind searches will fully sample the
Zone of Avoidance; first, a program of short (5-minute) integrations
has been searching for \hi emission from large, or particularly
nearby, hidden galaxies; a program of longer ($12 \times 5$-minutes)
observations has now been initiated looking for smaller, or more
distant systems.  Both blind searches involve observing on a honeycomb
grid with lattice points separated by $0\fdg35$; because the FWHM beam
of the 25-m telescope operating at a wavelength of 21 cm is 36
arcminutes, the honeycomb grid is fully sampling the beam.  The blind
searches are covering the regions of full optical obscuration at
$|b|<5\degr$, over most of the accessible range in longitude.  The
initial period of the survey has been directed toward the longitude
range in the second quadrant where the Supergalactic equator crosses
the Zone of Avoidance and where a large population of galaxies is
expected a priori.  (Most of the galaxies -- all of them previously
known ones -- identified as contaminating nuisances in the
Leiden/Dwingeloo survey of \hi in the Milky Way occur in a vertical
swath crossing the equator of the Milky Way near the center of the
second longitude quadrant: see e.g. figure 4.4 in Hartmann, 1994, or
figure 5 in Burton \& Hartmann, 1994.)  Although the blind searches
are the principal facets of the strategy, a limited directed search is
being carried out involving longer integrations toward candidate
objects selected from optical and infrared datasets.  Integration
times of an hour per spectrum yield sensitivities of about 0.01 K over
the range $0 < v < +4000$ km s$^{-1}$.

      Many problems can plague single-dish 21-cm spectra.  In
particular, the expected emission signature of an external galaxy can
be mimicked by interference signals of unusual type, i.e. not
appearing as a single-channel spike or as a ${\sin (x)}/x$ ringing
(see figure 2.13 of Hartmann, 1994); the signal from a low-velocity
galaxy like either Dw1 or Dw2, or like Maffei 2, might be at least
partly blended with that from conventionally-behaving gas associated
with the Milky Way, or it might resemble that of a (very)
high-velocity cloud associated with the Milky Way; the signal might be
confused with edge-of-bandpass effects; or the signal might be
contaminated by enhanced system temperatures due to the appearance of
a Milky Way \hii region, a plethora of which occur at the latitudes
being searched (see e.g. Lockman's 1989 catalog).  Each of these
problems could be dealt with at the 25-meter telescope itself, but an
important aspect of the strategy of the Dwingeloo Obscured Galaxies
Survey is the intention to seek rapid independent confirmation of
suspected detections by using the Westerbork Synthesis Radio Telescope
in the ``snapshot'' mode.  Thus the detection spectrum of Dwingeloo 1
(see Kraan--Korteweg et al., 1994) was discussed amongst members of
the project team in a meeting in Dwingeloo on Thursday, August 4,
1994, and by the end of the day on Monday, August 8, a WSRT snapshot
was in hand confirming the detection and giving preliminary
information on the \hi position and flux.  By the end of that week,
optical and near- as well as far-infrared identifications of the newly
detected galaxy had also been obtained (see Kraan--Korteweg et al.
1994; Loan et al. 1995).  We discuss here full $4 \times 12$ hour
synthesis observations of Dwingeloo 1, and the discovery in these
observations of another galaxy, which we call Dwingeloo 2 and which is
probably a companion to Dw1.

\begin{figure*}[tp] 
\picplace{23.5cm}
\end{figure*}

\section{WSRT observing program}

     The Westerbork Synthesis Radio Telescope is an interferometer
comprising an array of 14 telescopes, each with a diameter of 25
meters.  The east-west alignment implies that the technique of
earth-rotation aperture synthesis requires an observation of 12 hours
duration to obtain a synthesized field.  The observations reported
here were carried out in five sessions during August and September,
1994, and resulted in a full synthesis of $4.4 \times 12$ hours
duration, including the snapshot material.  The four eastern, movable
telescopes were set at different positions during the different
12-hour periods, and the signals from the fixed-movable telescope
pairs were correlated; this provided a total of 152 different
baselines ranging in length from 36 to 2754 meters with a standard
increment of 18 meters.  As a result, the instrumental grating rings
caused by the constant increment in baseline length lie well beyond
the FWHM of the primary beam and do not interfere with extended
emission.

     The synthesized beam has an elliptical shape with dimensions
$\alpha\times\delta = 11.8\times13.8$ arcsec.  The FWHM size of the
primary beam of the WSRT when operated at a wavelength of 21 cm is
$37\farcm 6$, quite similar to that of the single-dish Dwingeloo
telescope.  A series of spectra made with the 25-m telescope at
neighboring pointings around the source later to be identified as
Dwingeloo 1 had given a rough indication of the direction towards
which the WSRT should be pointed for the initial ``snapshot''
observation, and the suspicion (correct, as it turned out) that the
smudge on the red Palomar Sky Survey plate (\#936) catalogued by Hau
et al. (1995) might be the counterpart to the \hi feature, led to
centering the WSRT on the direction $\alpha=02^h53^m02^s$,
$\delta=+58\degr42'36''$ (epoch 1950).  The central frequencies of the
WSRT receivers were tuned to correspond to an {\it LSR} velocity of
106 km s$^{-1}$.  The single-dish detection had shown the
double-horned signature characteristic of \hi emission from a spiral
galaxy, but the extent to which the lower-velocity horn was
contaminated by emission from the Milky Way was not clear until the
WSRT data confirmed that the central-frequency tuning of the receivers
was reasonably accurate.  A total bandwidth of 2.5 MHz was covered by
63 channels; during the observations an on-line uniform frequency
taper was applied, yielding a velocity resolution of 9.89 km s$^{-1}$.

\begin{table}[t]
\caption{Westerbork observing parameters for the Dwingeloo 1 field}
\begin{center}
\begin{tabular}{ll}
\hline
field center (1950.0)\phantom{xxx} $\alpha$           & 02$^h$53$^m$02$^s$
  \\
\phantom{field center (1950.0)}\phantom{xxx} $\delta$ & 58$^{\circ}$42$'$36$''$
  \\
dates of observations                                 & 8 \& 12 Aug.\ 1994
  \\
                                                      & 6, 26 \& 29 Sept.\ 1994
  \\
length of observations                                & 4.4$\times$12 hours
  \\
number of different baselines                         & 152
  \\
baselines (min--max--incr)                            & 36--2754--18 meters
  \\
FWHM of synthesized beam ($\alpha \times \delta$)     & 11.8 $\times$ 13.8
arcsec \\
radius of first grating ring ($\alpha \times \delta$) & 40.1 $\times$ 47.0
arcmin \\
FWHM of primary beam                                  & 37.6 arcmin
  \\
rms noise in one channel map                          & 4.8 K
  \\
central frequency                                     & 1420.01 MHz
  \\
central velocity ({\it LSR})                          & 106 km s$^{-1}$
  \\
bandwidth                                             & 2.5 MHz
  \\
number of channels                                    & 63
  \\
channel separation                                    & 8.24 km s$^{-1}$
  \\
velocity resolution                                   & 9.89 km s$^{-1}$
  \\
K$\Rightarrow$mJy conversion,                         &
  \\
\phantom{xxx} equivalent of 1 mJy/beam                & 4.24 K
  \\
\hline
\end{tabular}
\end{center}
\label{WSRTobs}
\end{table}

     The interferometric $UV$-plane data were calibrated, flagged if
necessary, and Fourier transformed with the {\it NEWSTAR} software
package.  Before Fourier transforming, the $UV$-plane data were
convolved with an {\it expsinc} function to reduce aliasing effects,
and a Gaussian baseline taper was applied to suppress the sidelobes of
the synthesized beam.  Furthermore, the $UV$-data were weighted
according to the local density of data points in the $UV$-plane.  Due
to decreasing sensitivity near the edges of the bandpass, the first 3
as well as the last 7 channels were discarded.  This procedure
resulted in a data cube containing $1024 \times 1024 \times 53$
pixels, covering a field of angular dimensions $1\fdg42 \times
1\fdg66$ with a synthesized beam of FWHM of $11\farcs8$ in $\alpha$
and $13\farcs8$ in $\delta$.  The total velocity coverage was 428.7 km
s$^{-1}$.  Application of the various tapers and convolution functions
yielded an rms noise in the channel maps of 1.2 mJy per beam.  A set
of 5 maps with the antenna patterns at different frequencies was
constructed for use when applying the {\it CLEAN} algorithm to the
channel maps.

\begin{figure*}[t] 
\caption[2]{Global kinematic and spatial properties of the \hi
distribution in Dwingeloo 1.  Panel {\bf $a)$} shows, for comparison with
the global \hi situation, a greyscale representation of the INT
$I$-band image (see Loan et al. 1995).  The size and orientation of
the optical image is represented by the ellipse in panel {\bf $b)$}, which
shows the {\it CLEAN}ed continuum map, with contours drawn at levels
of $-$1.2, $-$0.58, 0.58 (the 2-$\sigma$ level), 1.2, 2.3, 4.6, 9.3,
and 19 K.  Panel {\bf $c)$} shows the global \hi line profile in mJy,
corrected for the sensitivity of the primary beam; the vertical arrow
shows the systemic velocity of Dw1, 110.3 km s$^{-1}$, as derived from
the velocity-field data.  Panel {\bf $d)$} shows the deprojected \hi surface
density in units of M$_\odot$ pc$^{-2}$, as determined in tilted,
rotating rings separately for the approaching and receding sides of
Dw1; the full-drawn line corresponds to the average of the surface
densities measured on the two sides; the vertical arrow shows the
angular extent of the semi-major axis of the optical image.  The
panels labelled {\bf $e)$} show position-velocity maps of the \hi emission
observed, at the indicated position angles, along the kinematic major
axis (left) and along the kinematic minor axis (right) of Dw1.
Contours in the major-axis position-velocity map correspond to
$-$0.36, $-$0.18 (the 2-$\sigma$ level), 0.18, 0.36, 0.73, 1.1, 1.5,
2.2, 2.9, and 3.6 $\times 10^{20}$ \hi atoms cm$^{-2}$; in the
minor-axis map, the contours correspond to $-$0.42, $-$0.21, 0.21
(2$\sigma$), 0.42, 0.85, 1.3, 1.7, 2.5, and 3.4 $\times 10^{20}$ \hi
atoms cm$^{-2}$. The dots follow the rotation curve as projected onto
this position-velocity maps.  In both panels, the horizontal dashed
line gives the systemic velocity while the vertical one corresponds to
the location of the center of rotation; the Milky Way contributes the
emission contaminating the lower-velocity portion of each of these
panels. The cross in the lower-lefthand corner of each of the
position-velocity slices gives the spatial and kinematic resolution of
the WSRT data.

     Panel {\bf $g)$} shows the observed velocity field determined for Dw1
by Gaussian fitting to individual \hi spectra.  The lighter greyscale
and the black contours correspond to the approaching side; the darker
greyscale and the white contours, to the receding.  The heavily-drawn
black contour follows the systemic velocity.  The interval between
contour lines is 20 km s$^{-1}$, measured from the systemic velocity
of 110.3 km s$^{-1}$.  Panel {\bf $i)$} shows the modelled velocity field as
a ``spider diagram'' derived by fitting the observed field with tilted
rings in which the \hi describes circular rotation; the isovelocity
contours in this panel are drawn as in {\bf $g)$}.  Panel {\bf $j)$} shows the
difference between the observed and modelled velocity fields; the
contours give the velocity residuals at levels of $-$10, $-$5, 5, and
10 km s$^{-1}$.  Panel {\bf {\bf $h)$}} shows the distribution of the
total \hi column density across Dw1.  As indicated in the text, the
noise level varies across this map.  Greyscaled pixels in the
total-\hi map all have a signal-to-noise ratio $\ge 1.5$; the contours
correspond to 1.8 (mean $S/N=3$), 3.0, 6.1, 9.1, 12, and 18 $\times
10^{20}$ \hi atoms cm$^{-2}$.  The three panels in {\bf $f)$} show the
radial variations of, respectively beginning with the upper of the
three, the inclination of tilted rings fit to Dw1, the position angle
of these rings, and the rotation velocity.  The rotation velocity was
determined separately for the approaching and receding sides, as
indicated.  The full-drawn line in the rotation-curve panel does not
correspond to the average of these two solutions, but to a separate
solution incorporating complete rings.  The filled circle in the two
upper panels and in the one on the lower left indicates the FWHM of
the synthesized WSRT beam.}
\end{figure*}

     Further reduction and analysis of the data utilized the Groningen
Image Processing SYstem ({\it GIPSY}), following practices described
by Verheijen (1996).  First, the data cube was smoothed to spatial
resolutions of 30 and 60 arcseconds in order to obtain higher
sensitivity to extended emission at low column densities.  These
smoothed data cubes showed that \hi emission from Dw1 covers the
velocity interval extending from 7 to 221 km s$^{-1}$; the emission
from Dw2 extends from 32 to 156 km s$^{-1}$.  Because the channels at
velocities below about 25 km s$^{-1}$ are contaminated by \hi emission
from the Milky Way, a map of the 21-cm {\it continuum} emission was
made by averaging 10 channels at velocities from 229.7 to 303.9 km
s$^{-1}$, which were judged to be free of any \hi emission either from
the Milky Way or from the galaxies in question.  The rms noise in the
resulting high-resolution continuum map is 0.37 mJy/beam.  The
continuum map was subtracted from each channel map in the data cube.
The continuum-subtracted channel maps which do contain \hi emission
from Dw1 and Dw2, as well as the continuum map itself, were {\it
CLEAN}ed by applying that algorithm in the standard manner (see
H\"ogbom 1974).  The map areas to be cleaned were determined by the
2-$\sigma$ contour in the un-{\it CLEAN}ed map at $60'' \times 60''$
resolution and extended by 1 arcmin in order to account for possible
emission suppressed by the sidelobes below the formal 2-$\sigma$
level.  Note that these areas vary from channel to channel due to the
rotation of both galaxies.  The selected areas were {\it CLEAN}ed down
to the 0.3-$\sigma$ level.  The {\it CLEAN} components were restored
by Gaussian beams of FWHM dimensions $11'' \times 13''$, $30'' \times
30''$, and $60'' \times 60''$, respectively.  The rms noise in the
continuum-subtracted and {\it CLEAN}ed channel maps is 1.2, 1.6, and
2.3 mJy/beam for the maps at high, medium, and low resolutions,
respectively.  After {\it CLEAN}ing, the map areas containing \hi
emission were redefined on the basis of the 2-$\sigma$ level in the
{\it CLEAN}ed maps at $60''\times 60''$ resolution.  The maps at $30''
\times 30''$ resolution give the best compromise between resolution
and sensitivity and are thus the ones considered further here.

     The global \hi profiles of Dw1 and Dw2 were determined by
summation of the total signal, primary-beam-corrected, in those areas.
Taking the primary beam attenuation into account, the sum of the two
global \hi profiles, one for Dw1 and one for Dw2, corresponds very
well to the single-dish detection spectrum measured toward Dw1 on the
Dwingeloo 25-m telescope by Kraan--Korteweg et al. (1994), as well as
with the spectrum observed subsequently by Huchtmeier et al. (1995)
using the Effelsberg 100-m telescope .

     Relevant parameters of the WSRT observing program are summarized
in Table 1.

\begin{figure*}[tp] 
\picplace{23.5cm}
\end{figure*}

\begin{figure*}[tp] 
\picplace{23.5cm}
\end{figure*}

\section{\hi in Dwingeloo 1}

\begin{figure*}[t] 
\caption[]{Mosaic of channel maps showing the \hi emission from
Dwingeloo 2, after subtraction of the background continuum radiation
and after application of the {\it CLEAN} algorithm, but uncorrected
for primary-beam attenuation.  For each channel map the appropriate
central {\it LSR} velocity is separately indicated; a cross indicates
the position of the dynamical center.  The lower-righthand panel shows
the (un-{\it CLEAN}ed) continuum radiation from the field, which was
subtracted from the individual channel maps.  Contour lines in the
channel maps are drawn as in Figure 1.  Contour lines in the continuum
map correspond to $-$1.3, $-$0.62, 0.62 (the 2-$\sigma$ level), 1.3,
2.5, 5.0, 7.5, 10, 13, and 15 K; the dashed-line contours represent
the negative values.}
\end{figure*}

     Fig. 1 is a mosaic of channel maps showing the \hi emission
from Dwingeloo 1; the panels represent alternate channels, each
corresponding to the emission over an interval 8.2 km s$^{-1}$ wide
centered at the indicated velocity.   The first five panels in the
series show contamination from \hi emission from the Milky Way.  The
last panel shows the subtracted continuum map.  \hi emission from Dw1
is evident in each of the displayed panels representing channels
centered between 15 and 213 km s$^{-1}$.  The emission shows the
``spider-diagram'' isovelocity characteristics which are the signature
of \hi in an inclined spiral galaxy.  The ellipse drawn in the
upper-lefthand panel of Fig. 1 represents schematically the
orientation and size of the optical image of Dw1 observed in the
$I$-band with the Isaac Newton Telescope on La Palma (see
Kraan--Korteweg et al. 1995; Loan et al. 1995); the image itself is
reproduced in Fig. 2a.  The spatial extent of the Dw1 \hi channel-map
slices is clearly larger than the extent of the optical-image ellipse.
Although such a situation is common for spiral galaxies, we note that
the optical obscuration at $b=-0\fdg11$ is so severe that the outer
boundary of the $I$-band image is probably not a quantitative measure
of the actual extended stellar dimensions of the system.

     Fig. 2 shows the principal large-scale kinematic and spatial
properties of the neutral hydrogen distribution in Dwingeloo 1.  The
global \hi profile shown in Fig. 2c was obtained by integrating the
individual continuum-subtracted channel maps, giving the \hi flux
density as a function of radial velocity.  The global profile shows
the two-horned signature characteristic of an inclined spiral galaxy,
with the flux enhanced at the edges of the profile by velocity
crowding.  The arrow in the global-profile panel indicates the
systemic velocity of $110.3\pm0.4$ km s$^{-1}$ as derived from the
velocity field of Dw1.  The radial velocity of Dw1 with respect to the
center of our Galaxy is $v_G = 110+220\cos (b) \sin (l) = 257$ km
s$^{-1}$, assuming a circular velocity of the LSR of 220 km s$^{-1}$.

     A map of the spatial distribution of the total integrated \hi
column density in Dw1 is shown in Fig. 2h.  This total-\hi map was
constructed by adding all the channel maps, after having set the
pixels outside the defined areas to zero.  Because the number of added
channels thus varies from pixel to pixel, the noise in the total-\hi
map varies as well from pixel to pixel, independent of the strength of
the signal.

     The panels of Fig. 2e show position-velocity diagrams
corresponding, respectively, to the major axis of the total-\hi map,
determined at a position angle of $112\degr$, and to the minor axis of
the total-\hi map, at a position angle of $202\degr$.

     The observed (projected) velocity field of Dw1 shown in Fig. 2g
was determined by fitting single Gaussians to each velocity profile
through the data cube.  Only those Gaussian solutions were accepted
which met the three-fold criterion of returning a peak inside the area
defining the \hi emission, of returning a peak amplitude greater than
3 times the noise in the profile in question, and of yielding a peak
velocity with an error less than a third of the velocity resolution of
the WSRT data.

     The kinematic and spatial characteristics of Dw1 mentioned
above are indicative of a well-behaved system.  Such behavior
justifies further analysis based on a model of tilted, rotating
rings.  This analysis technique has been described by Begeman
(1989).  Fig. 2i shows the results of modelling the observed velocity
field with such tilted rings.  The width of each ring was held
constant at value equal to 2/3 of the size of the major axis of the
synthesized WSRT beam.

     The ring fitting involved three steps.  First, the systemic
velocity, the center of rotation, and the rotation velocity were
modelled for each ring keeping the inclination and position angle
fixed at values derived from fitting ellipses to the total \hi map.
No significant trend in either systemic velocity or in rotation center
as a function of radius was detected; in the subsequent steps, the
rings were therefore treated as spatially and kinematically
concentric.  Second, the rotation velocity, the position angle, and
the inclination were fit for each ring keeping the systemic velocity
and center fixed at the previously-determined average values.  No
significant trend of inclination from the mean value of $51\degr$
(measured such that an edge-on galaxy is inclined $90\degr$) was
detected as a function of radius which might be taken as indication of
a morphological property of the galaxy such as warping; some localized
wiggles in the parameters might be due to the influence of spiral
structure.  A slight change of position angle detected near the center
of Dw1 may well be caused by non-circular motion due to the presence
of the bar evident on infrared and optical images of the galaxy (see
Kraan--Korteweg et al. 1994; Loan et al. 1995).  Finally, the rotation
velocity and the position angle were fit for each ring, keeping all
other parameters fixed.  This last step was done in half rings,
separately for the receding and for the approaching sides of Dw1, in
order to investigate the degree of symmetry of the rotating disk.  The
substantial symmetry of the system is indicated by the two panels of
Fig. 2e, where the mean rotation curve is projected on the
position-velocity diagrams.

    The orientations of the rings as derived by the tilted-ring model
were used to derive a radial \hi profile by averaging the \hi column
density from the total \hi map in annuli corresponding to the
projected rings.  The width of the annuli was set to 2/3 of the major
axis of the synthesized WSRT beam.  Fig. 2d shows the variation of the
deprojected \hi surface density in solar-mass units with angular
distance from the galactic center, plotted separately for the receding
and the approaching sides of Dw1.  The vertical arrow indicates the
angular extent of the semi-major axis of the $I$-band image.

     Finally, Fig. 2b shows the 21-cm continuum emission from Dw1.
This emission, which is contributed from the entire extended disk, has
an integrated flux of $48.2 \pm 7.5$ mJy and suggests on-going star
formation in the galaxy.

\begin{figure*}[tp] 
\picplace{23.5cm}
\end{figure*}

\section{\hi in the newly discovered galaxy Dwingeloo 2}

     The Westerbork observations resulted in the discovery of a small
galaxy, located some 21 arcminutes to the west-northwest of Dw1 and
quite likely its companion.  The \hi signal from this galaxy, which we
call Dwingeloo 2, would have been heavily blended in the single-dish
detection spectrum of Dw1.  The signals from the two systems are
centered less than 20 km s$^{-1}$ apart in velocity, and their angular
separation is less than the FWHM of the beam of the Dwingeloo 25-m
telescope.

     We were unable to find Dwingeloo 2 on the POSS I plates, but
inspection of the red POSS II plate revealed a feature of low surface
brightness, elongated like the WSRT \hi image. C. Aspin, and R.
Tilanus have detected the stellar component of Dw2 in observations
made using the IRCAM3 (see Puxley et al. 1994) camera operating at a
wavelength of $2.2\mu$m on the UKIRT facility on Mauna Kea.  They have
kindly provided the $K$-band image reproduced in Fig. 4a.  This image
reaches a depth of $m_K=20$; although the signal from the stellar
component of Dw2 is weak, it has evidently been detected as a smudge
inclined on the sky in the same manner as the gaseous component and as
the POSS II feature.  (We note that the possibilities now becoming
available of deep imaging at $2.2\mu$m, as well as of surveying at
this wavelength, will likely lead to upward revisions of the sizes and
stellar masses of heavily obscured galaxies, as well as to the
discovery of additional ones.)

     Fig. 3 is a mosaic showing the \hi emission from Dw2 in
individual channel maps, each centered at the indicated velocity.
Despite the decreased sensitivity (40\% of maximum) of the primary
beam at the off-axis position of Dw2, the small galaxy contributes
easily-detected signals to the panels between 32 and 155 km s$^{-1}$.
The ``spider-diagram'' signature so pronounced in Dw1 does not
characterize the Dw2 channel maps:  the \hi disk of Dw2 apparently
rotates as a solid body.  We note, however, that the \hi distribution
in each channel map is not quite perpendicular to the overall
kinematic major axis as would be expected for an unperturbed disk
rotating as a solid body.

     Fig. 4 shows the principal kinematic and spatial global
properties of the \hi distribution in Dw2; these properties were
extracted from the WSRT data in the same manner as described for Dw1.
The global \hi profile shown in Fig. 4c does not show the two-horned
signature, because of the absence of the velocity-crowding which would
follow from an extensive region spanned by an essentially flat
rotation curve.   The systemic velocity of Dw2 as derived from the
velocity field is $94.0\pm1.5$ km s$^{-1}$ and is indicated by the
arrow in the global-profile panel.  The radial velocity of Dw2 with
respect to the center of our Galaxy is $v_G = 94 + 220\cos (b) \sin
(l) = 241$ km s$^{-1}$.

     A map of the spatial distribution of the total integrated \hi
column density in Dw2 is shown in Fig. 4h.  The total-\hi distribution
is rather irregular and thus poorly suited to estimating the
inclination of Dw2.

     The projected velocity field shown in Fig. 4g was determined by
fitting single Gaussians to each velocity profile through the data
cube, in a procedure identical to that followed for the Dw1 data.  The
observed velocity field seems to be twisted, suggesting a dominant
central bar or a warp in the outer regions.  Modeling the velocity
field of Dw2 by tilted concentric rings, assuming circular motions and
thus ignoring a possible bar potential, shows a rather constant
inclination angle of $69\degr$ but a clear change in position angle
from $294\degr$ in the inner regions to $271\degr$ in the outer ones.
The results of the tilted-ring fit are shown in the three panels of
Fig. 4f.  Note that the kinematic position angles do not reflect the
position angle of the overall \hi distribution; this situation favors
the idea of a bar potential over that of a warp.

     The left-hand panel of Fig. 4e shows a position-velocity diagram
at the position angle of the kinematic major axis in the inner region
of Dw2; the right-hand panel shows the position-velocity slice along
the kinematic minor axis in the inner region.  Both diagrams are
centered on the derived dynamical center of Dw2.  The dots represent
the rotation curve projected on to these slices.

     Fig. 4d shows the deprojected \hi surface density, in units of
M$_\odot$ pc$^{-2}$, plotted separately for the approaching and
receding sides of Dw2.  The integrations were done in ellipses
prescribed by the results from the tilted-ring modelling.  Although
the central \hi surface densities in Dw 1 and in Dw2 are quite
similar, the densities in Dw2 fall rapidly with distance from its
center.

\begin{table}[t]
\caption{Rotation curves (deprojected for inclination) for Dwingeloo
1 and Dwingeloo 2, derived from least-squares modelling of concentric,
rotating, tilted rings}
\begin{center}
\begin{tabular}{rcccccc}
    & &\multicolumn{2}{c}{Dwingeloo 1}& &\multicolumn{2}{c}{Dwingeloo 2}\\
\hline
 R  & & v$_{\mbox{rot}}$ &    P.A.    & & v$_{\mbox{rot}}$ &   P.A.    \\
($''$)& &(km$\;$s$^{-1}$)& ($^\circ$) & & (km$\;$s$^{-1}$) & ($^\circ$) \\
\hline
 20 & &  20.9$\pm$1.6 & 135.2$\pm$5.5 & & 19.5$\pm$2.6 & 296.7$\pm$6.0 \\
 40 & &  42.3$\pm$1.1 & 127.3$\pm$2.1 & & 28.4$\pm$1.2 & 297.9$\pm$2.4 \\
 60 & &  67.8$\pm$1.0 & 115.4$\pm$1.1 & & 33.5$\pm$1.3 & 291.4$\pm$1.9 \\
 80 & &  83.8$\pm$0.7 & 114.3$\pm$0.6 & & 37.9$\pm$1.8 & 281.8$\pm$2.7 \\
100 & &  90.9$\pm$0.6 & 112.0$\pm$0.5 & & 43.5$\pm$0.9 & 274.8$\pm$1.3 \\
120 & &  94.1$\pm$0.5 & 111.3$\pm$0.4 & & 48.0$\pm$0.9 & 269.5$\pm$1.1 \\
140 & &  97.9$\pm$0.5 & 113.2$\pm$0.4 & & 51.6$\pm$1.0 & 271.4$\pm$1.1 \\
160 & & 101.6$\pm$0.4 & 114.2$\pm$0.3 & &              &               \\
180 & & 104.9$\pm$0.5 & 113.8$\pm$0.4 & &              &               \\
200 & & 108.3$\pm$0.5 & 113.0$\pm$0.4 & &              &               \\
220 & & 110.7$\pm$0.4 & 112.4$\pm$0.3 & &              &               \\
240 & & 112.4$\pm$0.4 & 111.9$\pm$0.3 & &              &               \\
260 & & 113.5$\pm$0.4 & 111.6$\pm$0.3 & &              &               \\
280 & & 113.6$\pm$0.4 & 111.8$\pm$0.3 & &              &               \\
300 & & 112.7$\pm$0.5 & 112.3$\pm$0.3 & &              &               \\
320 & & 113.9$\pm$0.5 & 113.0$\pm$0.3 & &              &               \\
340 & & 115.4$\pm$0.4 & 112.4$\pm$0.3 & &              &               \\
360 & & 114.2$\pm$0.5 & 111.7$\pm$0.3 & &              &               \\
380 & & 112.3$\pm$0.5 & 112.0$\pm$0.3 & &              &               \\
400 & & 110.6$\pm$0.6 & 111.2$\pm$0.4 & &              &               \\
420 & & 110.2$\pm$0.6 & 112.0$\pm$0.4 & &              &               \\
440 & & 109.0$\pm$1.1 & 114.9$\pm$1.0 & &              &               \\
\hline
\end{tabular}
\end{center}
\label{Dw12RC}
\end{table}

\section{Concluding remarks on the intrinsic properties of Dwingeloo 1
and Dwingeloo 2}

\begin{figure*} 
\caption[]{Global kinematic and spatial properties of the \hi
distribution in Dwingeloo 2.  Panel {\bf $a)$} shows, for comparison with
the global \hi situation, a greyscale representation of the UKIRT
$K$-band image kindly provided by R. Tilanus and C. Aspin.  Panel {\bf
$b)$} shows the {\it CLEAN}ed continuum map, with contours drawn at
levels of $-$1.2, $-$0.58, 0.58 (the 2-$\sigma$ level), 1.2, 2.3, 4.6,
9.3, and 19 K.  Panel {\bf $c)$} shows the global \hi line profile in
mJy, corrected for the sensitivity of the primary beam; the vertical
arrow shows the systemic velocity of Dw2, 94.0 km s$^{-1}$, as derived
from the velocity-field data.  Panel {\bf $d)$} shows the deprojected
\hi surface density in units of M$_\odot$ pc$^{-2}$, as determined in
tilted rings separately for the approaching and receding sides of Dw2;
the full-drawn line corresponds to the average of the surface
densities measured on the two sides.  The panels labelled {\bf $e)$}
show position-velocity maps of the \hi emission observed, at the
indicated position angles, along the major axis (left) and along the
minor axis (right) of Dw2.  Contours in the major-axis
position-velocity map correspond to $-$0.21, 0.21 (the 2-$\sigma$
level), 0.42, and 0.85 $\times 10^{20}$ \hi atoms cm$^{-2}$; in the
minor-axis map, the contours correspond to $-$0.20, 0.20 (2$\sigma$),
0.40, 0.81, and 1.2 $\times 10^{20}$ \hi atoms cm$^{-2}$. The dots
follow the rotation curve as projected onto this position-velocity
maps.  In both panels, the horizontal dashed line gives the systemic
velocity while the vertical one corresponds to the location of the
center of rotation. The cross in the lower-lefthand corner of each of
the position-velocity slices gives the spatial and kinematic
resolution of the WSRT data on Dw2.

     Panel {\bf $g)$} shows the observed velocity field determined for Dw2
by Gaussian fitting to individual \hi spectra.  The lighter greyscale
and the black contours correspond to the approaching side; the darker
greyscale and the white contours, to the receding.  The heavily-drawn
black contour follows the systemic velocity.  The interval between
contour lines is 15 km s$^{-1}$, measured from the systemic velocity
of 94.0 km s$^{-1}$.  Panel {\bf $i)$} shows the modelled velocity field as
a ``spider diagram'' derived by fitting the observed field with tilted
rings in which the \hi describes circular rotation; the isovelocity
contours in this panel are drawn as in {\bf $g)$}.  Panel {\bf $j)$} shows the
difference between the observed and modelled velocity fields; the
contours give the velocity residuals at levels of $-$10, $-$5, 5, and
10 km s$^{-1}$.  Panel {\bf {\bf $h)$}} shows the distribution of the total \hi
column density across Dw2.  As indicated in the text, the noise level
varies across this map.  Greyscaled pixels in the total-\hi map all
have a signal-to-noise ratio $\ge 1.5$; the contours correspond to 1.3
(mean $S/N=3$), 2.2, 4.2, 6.3, 8.4, and 13 $\times 10^{20}$ \hi atoms
cm$^{-2}$.  The three panels in {\bf $f)$} show the radial variations of,
respectively beginning with the upper of the three, the inclination of
tilted rings fit to Dw2, the position angle of these rings, and the
rotation velocity.  The shaded circle in the lower left of the two
upper panels and in the lowermost left-hand panel indicates the FWHM
of the synthesized WSRT beam.}
\end{figure*}

     Figs. 5 and 6 place Dw1 and Dw2 together in their relative
position and orientation on the sky, with Fig. 5 showing the total \hi
maps for the two galaxies and Fig. 6 their velocity fields.

     The assignment of Dw1 to the class of barred spiral galaxies is
most clearly required by the appearance of the UKIRT $K$-band image
obtained on August 10, 1994, by R. Tilanus, C. Aspin, and T. Geballe.
(This image was printed in the IAU General Assembly Newsletter
``Sidereal Times'', \#3, August 18, 1994.)  The $K$-band data show the
unmistakable signature of a bar, oriented along the major axis of the
\hi distribution as determined from the WSRT data, and extending to
about 1.5 arcminutes from the center of the galaxy.  The INT and Wise
Observatory $I$-band images confirm this assignment (see Loan et al.
1995).  Both the $K$- and $I$-band images show spiral arms emanating
from the ends of the stellar bar, and winding over some $180\degr$ in
azimuth.  The superposition shown in Fig. 7 of the WSRT total-\hi map
overlayed on the $I$-band image shows that the optically-traced
structure is continued to larger radii in the \hi distribution.  The
residual velocity-field map shown in Fig. 2j suggests that the spiral
structure at larger radii perturbs the kinematics of Dw1 in a manner
not uncommon for such galaxies.  The plot of \hi surface density
against distance from the center shown in Fig. 2d shows a relative
deficiency of \hi emission from the region of Dw1 occupied by the
stellar bar; such morphology is also not uncommon.

     Dw1 displays a quite simple morphology.  There is no indication
of a substantial lopsidedness in the \hi surface distribution, nor is
there any compelling indication of the kinematic or spatial
distortions which would be expected if the galaxy was severely warped,
or if it were undergoing an interaction with a close-by neighboring
system of comparable or higher mass.

\begin{figure*}[tp] 
\picplace{23.5cm}
\end{figure*}

\begin{figure*}[tp] 
\picplace{23.5cm}
\end{figure*}

     Dw2 is located approximately along the extension of the major
axis of Dw1.  Unlike the simplicity pertaining for the larger galaxy,
the kinematic and spatial properties of Dw2 do suggest that this
smaller system might be distorted by an interaction with the more
massive Dw1.  The isovelocity contours shown in Fig. 6 and in Fig. 4g
show a twist which might be associated with a gravitational distortion
of the smaller system by the larger one.  The panels in Fig. 4f
showing the variation of inclination and of position angle of the
fitted tilted rings additionally suggest a distortion of Dw2.  (The
panels in Fig. 2f showing the same information for Dw1 indicate no
significant variation of inclination or of position angle over
distance from the center of Dw1.)

\begin{figure*}[t] 
\caption[]{Distribution on the sky of \hi column densities measured in
Dwingeloo 1 and Dwingeloo 2.  The column densities are given by
the false-color coding as indicated by the color bar on the right-hand
side of the figure.  The circle in the lower-lefthand corner
represents the FWHM of the synthesized WSRT beam. Dw2 is located at an
angular separation from Dw1 of some 21$'$, approximately along an
extension of the major axis of Dw1.  If the two galaxies are located
at the same distance from the Sun, and if this distance is 3.8 Mpc
(see Loan et al. 1995; cf. Kraan--Korteweg et al. 1994), then the
angular separation corresponds to a linear distance of 24 kpc.}
\end{figure*}

\begin{figure*} 
\caption[]{Velocity fields revealed by \hi observations of Dwingeloo 1
and Dwingeloo 2.  The velocities are given by the false-color coding
as indicated by the color bar on the right-hand side of the figure.
The circle in the lower-lefthand corner represents the FWHM of the
synthesized WSRT beam.  Dw2 is located near the extended major axis of
Dw1.  There is some indication of a kinematic distortion in the
smaller system which tentatively suggests that Dw2 is a companion to
Dw1.}
\end{figure*}

\begin{table*}[t]
\caption{Intrinsic properties of Dwingeloo 1 and Dwingeloo 2 derived
from the WSRT observations}
\begin{center}
\begin{tabular}{lrll}
                                                           &
    & Dwingeloo 1          & Dwingeloo 2          \\
\hline
dynamical center (eq. coord.)\phantom{xxx} $\alpha_{1950}$           &
    & 02$^h$53$^m$01$^s$\negthinspace .6  &  02$^h$50$^m$21$^s$\negthinspace .7
\\
\phantom{dynamical center (eq. coord.)}\phantom{xxx} $\delta_{1950}$ &
    & 58$^d$42$'$43$''$    & 58$^d$48$'$07$''$    \\
 dynamical center (gal. coord.)\phantom{xxx} $l$           &
    &  $138\fdg52$         &  $138\fdg17$         \\
 \phantom{dynamical center (gal. coord.)}\phantom{xxx} $b$ &
    & $-0\fdg11$           & $-0\fdg19$           \\
 integrated HI flux                                        &
(Jy$\;$km$\;$s$^{-1}$) & 205.2 $\pm$ 1.2      &  30.8 $\pm$ 0.5      \\
 HI profile widths, 20\%                                   & (km$\;$s$^{-1}$)
    & 201.2 $\pm$ 0.4      & 116.6 $\pm$ 1.6      \\
 \phantom{HI profile widths, }50\%                         & (km$\;$s$^{-1}$)
    & 187.6 $\pm$ 0.6      &  99.9 $\pm$ 1.2      \\
 systemic velocity ({\it LSR})                             & (km$\;$s$^{-1}$)
    & 110.3 $\pm$ 0.4      & 94.0 $\pm$ 1.5       \\
 diameter of HI disk                                       & (arcmin)
    & 13.7                 & 6.4                  \\
 continuum flux at $\lambda 21$cm                          & (mJy)
    & 48.2 $\pm$ 7.5       & $<16^1$              \\
 inclination                                               & (degrees)
    & 51 $\pm$ 2           & 69 $\pm$ 3           \\
\hline
\multicolumn{4}{l}{$^1${\footnotesize 3-$\sigma$ limit for extended continuum
emission from Dw2}} \\
\end{tabular}
\end{center}
\label{Results}
\end{table*}

     Table 2 gives the rotation curves for Dwingeloo 1 and Dwingeloo
2, as derived from least-squares modelling of the observed velocity
fields by concentric, rotating, tilted rings; the values tabulated
have been corrected for the inclinations of the galaxies.  We note
that the \hi surface density of Dw2 drops below detectability in the
outer regions before reaching a regime of flat rotation.

     The suggested distortion of Dw2 by Dw1 remains tentative until
the linear distance between the two systems is better known.  The
matter of the distance to the two galaxies is, of course, important in
several regards.  It is not reasonable to use the Hubble-flow to gauge
the distance to such nearby systems.  Instead, Kraan--Korteweg et al.
(1994) estimated the distance to Dw1 at about 3 Mpc on the basis of
the blue Tully--Fisher relationship between the extinction-corrected
absolute magnitude and the inclination-corrected \hi linewidth.  The
galactic extinction was determined by Kraan--Korteweg et al. following
the precepts of Burstein \& Heiles (1982) and using the \hi column
density determined from the data of the Leiden/Dwingeloo survey of
Hartmann \& Burton (1995).  The principal uncertainties in this
estimate stem from the large scatter in the blue Tully--Fisher
relationship and in the unprecedented use of the Burstein--Heiles
relationship at the high values of $N_{\rm HI}$ found near to
$b=0\degr$.  It is not at all ruled out that the Galactic \hi spectra
so close to the equator are partially saturated (see Burton, 1993),
and if saturation does cause the extinction derived from the
integrated \hi to be underestimated, then the distance of 3 Mpc may
also be underestimated.  Loan et al. (1995) derived the distance to
Dw1 using the $I$- and $R$-band versions of the Tully--Fisher
relation; this has the advantage that corrections for Galactic and
internal extinctions are smaller in the near infrared than in the
blue.  Loan et al. choose as the most likely distance for Dw1 the
value 300 km s$^{-1}$, although they stress that the distance
uncertainties remain substantial.  Assuming that the value of the
Hubble constant is 80 km s${-1}$ Mpc$^{-1}$, we are led to adopt the
value of 3.8 Mpc as the distance to Dw1 in order to proceed with
estimating characteristic physical aspects of the galaxies.

     If Dw1 is at a distance of 3.8 Mpc then its \hi angular diameter
corresponds to a linear size of 15 kpc.  This estimate of the \hi
diameter may be compared with the dimension of the $K$-band bar as
projected at 3.8 Mpc distance, namely about 2.5 kpc, and with the
dimension of the IRAS 60-$\mu$m image (see Loan et al. 1995), namely
about 4 kpc.  The flat regime of the \hi rotation curve pertains
beyond about 4 kpc.  If Dw2 is at the same distance as Dw1, its \hi
diameter is 7 kpc.  The projected angular distance between Dw1 and Dw2
corresponds to a lower-limit on their linear separation of only 24
kpc.

     The total \hi mass harbored by the galaxies follows from
$2.36\times 10^5 D^2 I_{\rm HI}$, where $D$ is the distance in Mpc and
$I_{HI}$ is the integrated \hi flux.  For Dw1, $M_{\rm
HI}=7.0\times10^8 $M$_{\odot}$; for Dw2, $M_{\rm HI}=1.1\times10^8
$M$_{\odot}$.

\begin{figure*} 
\begin{center}
\picplace{10.5cm}
\end{center}
\end{figure*}

\begin{figure*} 
\begin{center}
\picplace{10.5cm}
\end{center}
\end{figure*}

\begin{figure*}
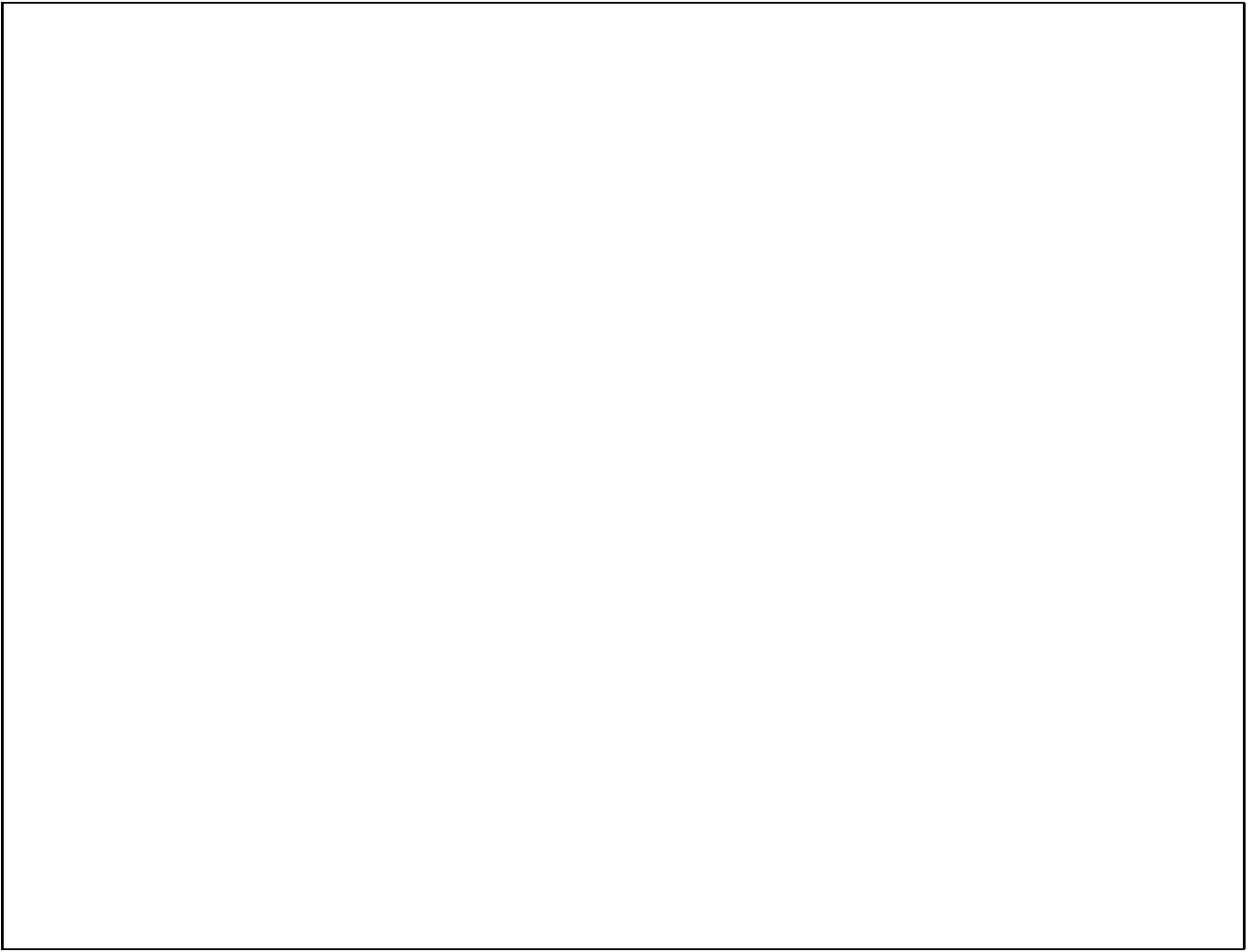
 
\begin{center}
\picplace{13.7cm}
\caption[]{Superposition of the WSRT map of the Dw1 \hi surface
densities, shown by contours, on the $I$-band false-color image from
the INT (Loan et al. 1995). The \hi surface density contours are drawn
at the same levels as those in Fig. 2b; regions of highest density are
encircled by the red contour.  The circle in the lower-lefthand corner
represents the FWHM of the synthesized WSRT beam.}
\end{center}
\end{figure*}

\begin{figure*} 
\begin{center}
\picplace{13.7cm}
\caption[]{Superposition of the WSRT map of the Dw2 \hi surface
densities, shown by contours, on the $K$-band false-color image from
the UKIRT IRCAM3 image kindly provided by R. Tilanus and C. Aspin.
The \hi surface density contours are drawn at the same levels as those
in Fig. 4b; regions of highest density are encircled by the red
contour.  The circle in the lower-lefthand corner represents the FWHM
of the synthesized WSRT beam.}
\end{center}
\end{figure*}

     The above values for the \hi and near- and far-infrared-image
diameters as well as for the \hi mass of Dw1 are consistent with what
would characterize a modest barred spiral galaxy; the values for Dw2
also seem plausible for the small companion.  The total mass (i.e.
including stellar as well as any dark matter) within a radius $r$ can
be estimated from $M_{tot}(<r) = (v_{rot})^2r/G$, where $v_{rot}$ is
the deprojected rotation speed.  For Dw1, $v_{rot}$ is about 113 km
s$^{-1}$.  Taking $r$ as the distance of the last measured point on
the rotation curve (see Table 2), namely at about $7\farcm5$ for Dw1
or at about $2\farcm5$ for Dw2, then the total mass to this distance
in Dw1 follows at the value $3.1 \times 10^{10}$ M$_\odot$ for Dw1 and
at the value $2.3 \times 10^9$ M$_\odot$ for Dw2.  The ratio
$M_{\rm HI}/M_{tot}$ would be about 0.02 for Dw1 and about 0.05 for
Dw2; these values seem plausible for a spiral and for a gas-rich
dwarf, respectively.

     It is interesting to compare the total mass of Dw1 with that of
the Milky Way as well as with that of Dw2; this comparison
follows simply from the squared ratio of rotation speeds in the outer
flat-rotation regimes, which is $M_{MW}/M_{Dw1}=(220/113)^2 = 3.8$ for
the ratio of total mass of the Milky Way compared to Dw1.  For Dw2,
where \hi has not been detected from a flat-rotation regime, we take
$v_{rot}$ at the maximum measured value of about 52 km s$^{-1}$.  This
value then yields a rough estimate of the ratio of $M_{Dw1}/M_{Dw2}
\sim (113/52)^2 = 4.7$ for the Dw1/Dw2 comparison of total masses.
These ratios give reasonable estimates of the relative masses within
the relevant radii, but do not give accurate estimates of the relative
{\it total} masses, because the true total extents are not yet known.
At the presumed distance and with the total mass estimated as
indicated, the mass of Dw1 would be comparable to that of M33.

     The distance estimate of 3.8 Mpc and the angular proximity of Dw1
\& 2 to the galaxies Maffei 1 \& 2 and IC342, make the conclusion
plausible that these galaxies are all members of the same grouping.
Recently McCall \& Buta (1995) discovered two dwarf companions to
Maffei 1; although these dwarfs would probably contribute little to
the overall mass budget of the group, their discovery additionally
motivates searches for galaxies in the vicinity.  We note that before
the detection of Dw1 there were several predictions, for different
reasons, concerning the likelihood of an undetected galaxy near Maffei
1 and 2.  Hurt et al. (1993) pointed out a number of morphological
distortions and other features evident in a deep $K$-band image of
Maffei 2 that were taken as indications of a recent interaction with a
companion galaxy.  Peebles (1990) concluded that the
motions of the Maffei group suggested the presence of a nearby galaxy,
as then undetected.  Whether or not the two Dwingeloo galaxies
exercise any significant dynamical influence on the motions of the
Maffei/IC342 group remains a matter for further consideration.

\begin{acknowledgements}
The Westerbork Synthesis Radio Telescope and the Dwingeloo 25-meter
telescope are operated by the Netherlands Foundation Research in
Astronomy with support from the Netherlands Organization for Scientific
Research (NWO).  While in Groningen the research of R.C.K.-K.  was made
possible by a fellowship from the Royal Netherlands Academy of Arts and
Sciences.  We are grateful to D.  Moorrees for continuing assistance
with the Dwingeloo Obscured Galaxies Survey observing program on the
25-m telescope, and to A.  Foley, H.  Kahlmann, and T.  Spoelstra for
the rapid scheduling and initial calibration of the WSRT observations.
We are grateful to R.  Tilanus and C.  Aspin for providing the UKIRT
$K$-band image of Dw2.
\end{acknowledgements}

{}

\end{document}